\documentclass{an}
\usepackage{times}
\usepackage{graph}

\begin{document}
\Pagespan{1}{4}
\Yearpublication{2007}
\Yearsubmission{2007}
\Month{}
\Volume{}
\Issue{}
\DOI{}
\title{ARTEMiS (Automated Robotic Terrestrial Exoplanet Microlensing Search) --
 A possible expert-system based cooperative effort to hunt for planets of Earth mass and below}
\author{M.~Dominik\inst{1}\fnmsep\thanks{Royal Society University Research Fellow}\fnmsep\thanks{Corresponding author:
  \email{md35@st-andrews.ac.uk}\newline}
K.~Horne\inst{1}, A.~Allan\inst{2}, N.J.~Rattenbury\inst{3},
Y.~Tsapras\inst{4}, C.~Snodgrass\inst{5}, M.F.~Bode\inst{4},
M.J.~Burgdorf\inst{4}, S.N.~Fraser\inst{4}, 
E.~Kerins\inst{3}, C.J.~Mottram\inst{4}, 
I.A.~Steele\inst{4}, R.A.~Street\inst{6}, P.J.~Wheatley\inst{7}, \L.~Wyrzykowski\inst{8,9}}
\institute{SUPA, University of St Andrews, School of 
Physics \& Astronomy, North Haugh, St Andrews, KY16 9SS, United Kingdom
\and School of Physics, University of Exeter, Stocker Road, Exeter EX4 4QL, United Kingdom
\and Jodrell Bank Observatory, Macclesfield, Cheshire, SK11 9DL, United Kingdom
\and Astrophysics Research Institute, Liverpool John Moores University, Twelve Quays House, Egerton Wharf, Birkenhead, CH41 1LD, United Kingdom
\and European Southern Observatory (ESO), Casilla 19001, Santiago de Chile, Chile
\and Las Cumbres Observatory Global Telescopes
Network, 6740B Cortona Dr, Goleta, CA 93117, United States of America
\and Department of Physics, University of Warwick, Coventry, CV4 7AL, United Kingdom
\and Institute of Astronomy, University of Cambridge, Madingley Road, Cambridge CB3 0HA, United Kingdom
\and Warsaw University Astronomical Observatory,
Al.\ Ujazdowskie 4, 00-478 Warszawa, Poland}
\received{} \accepted{}
\titlerunning{ARTEMiS -- An expert system to hunt for planets of Earth mass and below}
\authorrunning{M. Dominik et al.}
\publonline{}
\abstract{
The technique of gravitational microlensing is currently unique in its
ability to provide a sample of terrestrial exoplanets around both
Galactic disk and bulge stars, allowing to measure their abundance
and determine their distribution with respect to mass and orbital separation. Thus, valuable information for testing
models of planet formation and orbital migration
is gathered, constituting an
important piece in the puzzle for the existence of life forms
throughout the Universe. In order to achieve these goals in
reasonable time, a well-coordinated effort involving a network of either 2m
or 4 $\times$ 1m telescopes at each site is required. It could lead to the
first detection of an Earth-mass planet outside the Solar system, and even
planets less massive than Earth could be discovered. From April 2008, ARTEMiS (Automated Robotic Terrestrial Exoplanet
Microlensing Search) is planned to provide a platform for a three-step strategy of
survey, follow-up, and anomaly monitoring. As an expert system embedded in
eSTAR (e-Science Telescopes for Astronomical Research), ARTEMiS will give advice for follow-up based on a priority algorithm
that selects targets to be observed in order to maximize the expected number of planet detections, and will also alert on deviations from ordinary microlensing light curves by means of
the {\sc SIGNALMEN} anomaly detector. While the use of the VOEvent
(Virtual Observatory Event) protocol allows a direct interaction with the telescopes that are part of the HTN (Heterogeneous Telescope Networks) consortium, additional
interfaces provide means of communication with all existing microlensing campaigns that rely on human observers. The success of discovering a planet by microlensing critically depends on the availability of a telescope in a
suitable location at the right time, which can mean within
10~min. To encourage follow-up observations, microlensing campaigns are
therefore releasing photometric data in real time. On ongoing planetary anomalies, world-wide efforts are being undertaken to make sure that sufficient data are obtained, since there is no second chance. Real-time modelling offers the opportunity of live discovery of
extra-solar planets, thereby providing ``Science live to your home''.}

\maketitle

\section{Detecting planets by microlensing}
If an observed star in the Galactic bulge at distance $D_\mathrm{S}$ happens to be aligned with a foreground star with mass $M$ at distance $D_\mathrm{L}$ within an angle $\theta \la \theta_\mathrm{E}$, where
\begin{equation}
\theta_\mathrm{E} = \sqrt{\frac{4GM}{c^2}\,\left(D_\mathrm{L}^{-1}
-D_\mathrm{S}^{-1}\right)}
\end{equation}
is the {\em angular Einstein radius},
it exhibits a transient brightening (Einstein 1936) over about a month due to the bending
of its light by the gravitational field of the foreground star, 
constituting a {\em gravitational microlensing event}.

If moreover a planet orbiting the foreground 'lens' star happens to
be separated by an angle $\delta \sim \theta_\mathrm{E}$, it can reveal
its existence by means of a short distortion to the observed light curve
(Mao \& Paczy\'{n}ski 1991), depending on its mass lasting between several hours and several days. Besides three other current claims (Bond et al. 2004; Udalski et al. 2005; Gould et al. 2006), microlensing led to
the discovery of OGLE-2005-BLG-390Lb, the first cool rocky/icy exoplanet ever found (Beaulieu et al. 2006; Dominik, Horne \& Bode 2006).

Microlensing observations can provide samples of planets orbiting
stars in two distinct populations, namely the Galactic disk and the Galactic bulge, rather than facing a restriction to the Solar
neighbourhood. This will allow
an estimate of the abundance of planets in the Universe and provide a powerful
test of theoretical models of planet formation and orbital migration.

Given the properties of the two stellar populations, 
a typical microlensing event on an observed bulge star at $D_\mathrm{S}
\sim 8.5~\mbox{kpc}$ involves a lens star with $M \sim 0.3~M_{\odot}$ 
at $D_\mathrm{L} \sim 6.5~\mbox{kpc}$. This implies that
$\theta_\mathrm{E} \sim 350~\mu\mbox{as}$, so that microlensing is most sensitive to the detection of planets at a separation of 1 to 10 AU.
Since the respective orbital period by far exceeds the duration of the planetary signal, the latter reflects a snapshot of the planet at its current position.

\section{Microlensing observation campaigns}
With only about one in a million monitored stars being significantly 
brightened by the gravitational field of a foreground star
(Kiraga \& Paczy\'{n}ski 1994), 
the OGLE (Optical Gravitational Lensing Experiment) and 
MOA (Microlensing Observations in Astrophysics) surveys monitor more than
100 million stars on a daily basis, which results in 700-1000 microlensing events per year being alerted on-line while they are in progress (Udalski et al. 1992; Muraki et al. 1999; Bond et al. 2001; Udalski 2003). Their sampling is however insufficient
for detecting planets with masses significantly below that of Jupiter
(Snodgrass, Tsapras, \& Horne 2004).

The first microlensing follow-up network combining hourly sampling with
a round-the-clock coverage was established by PLANET (Probing Lensing
Anomalies Network) in 1995 (Albrow et al. 1998; Dominik et al. 2002).
While this network of 1m-class telescopes relies on human observers and
dedicated observing time, the demand of not only an immediate response, but also a flexible scheduling makes robotic telescopes ideally suited to carry out such an observing programme. Since 2004 -- and since 2005 in cooperation
with PLANET --, microlensing observations have been carried out with the
RoboNet-1.0 network of UK-built 2m robotic telescopes (Burgdorf et al. 2007), using a priority algorithm that selects those targets for which a detection of a planet is most likely to occur.
In contrast to PLANET/RoboNet, the MicroFUN team 
concentrate on a few quite promising events, with a network only
being activated on target-of-opportunity basis.

With the deployment of the {\sc SIGNALMEN} anomaly detector (Dominik et al. 2007), 
a combined effort of microlensing campaigns can realize a three-step strategy of survey, follow-up, and anomaly monitoring that allows for a substantial detection efficiency even for Earth-mass planets.

\section{Discovering planets of Earth mass and below}

\begin{figure*}
\begin{center}
\includegraphics[width=10.8cm]{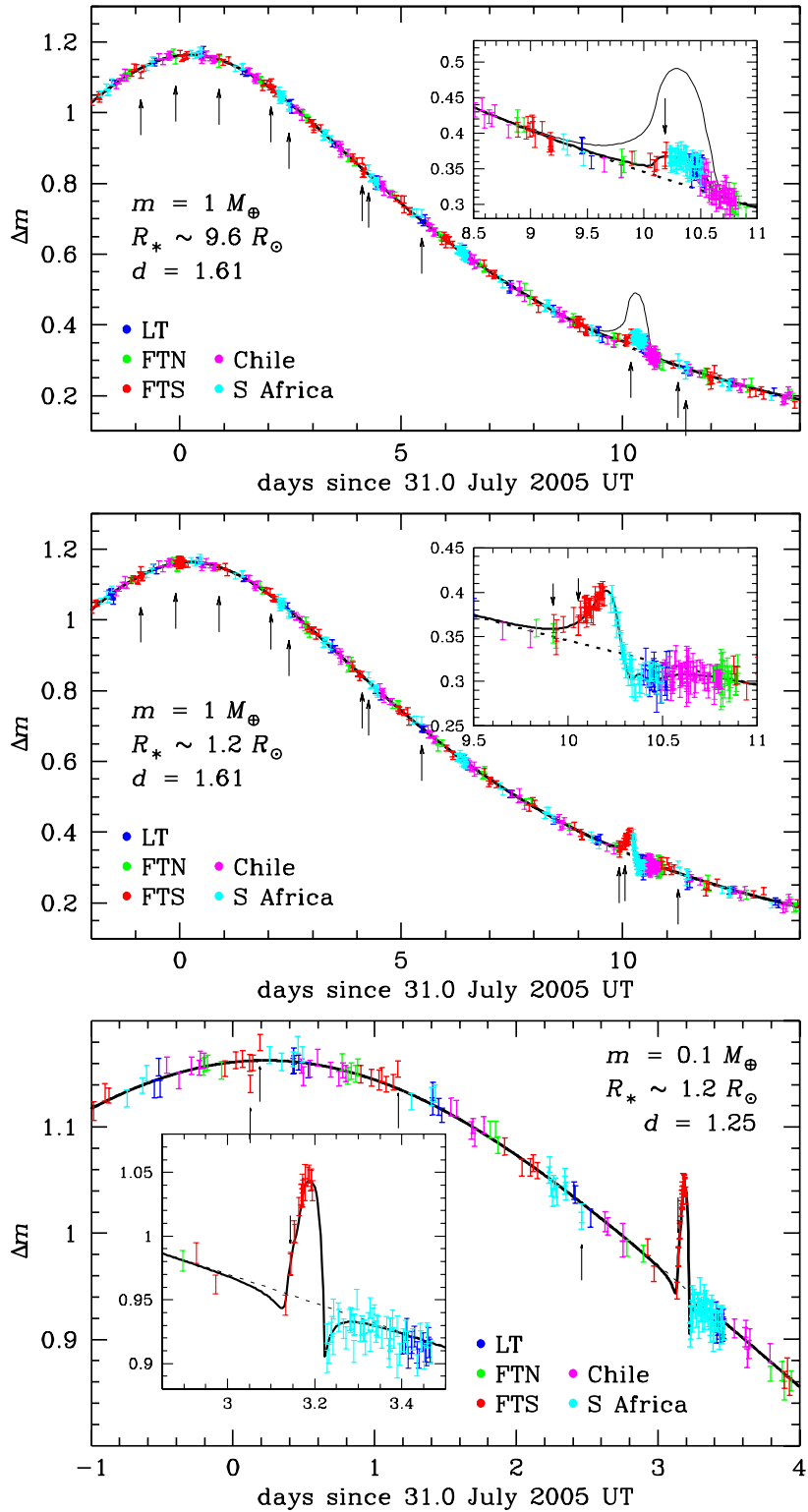}
\end{center}
\caption{The possible detection of planets of Earth mass or below (simulations) in three different configurations (where $d = \delta/\theta_\mathrm{E}$) with the robotic 2m-telescopes that constitute the 
RoboNet-1.0 network, namely the Liverpool Telescope (LT), the Faulkes Telescope North (FTN), and the Faulkes Telescope South (FTS), augmented
by two similar hypothetical telescopes located in Chile and
South Africa. Arrows indicate the epochs where the {\sc SIGNALMEN} anomaly
detector requested further observations. (top) 1-$M_\oplus$ planet in 
the same spot as OGLE-2005-BLG-390Lb, with the original model light curve
for the respective event also plotted; (middle) 1-$M_\oplus$ planet in
the same spot, but with a main-sequence source star; (bottom) 0.1-$M_\oplus$ planet at a closer distance ($d = 1.25$ instead of $d= 1.61$) to the lens star, and
a main-sequence source star. For the two latter cases, the orientation 
angle of the source trajectory with regard to the planet-star axis has
been modified from the OGLE-2005-BLG-390 model in order to produce a
deviation of the desired amplitude.}
\label{fig:SIGNALMENdetect}
\end{figure*}

\begin{figure*}
\begin{center}
\includegraphics[width=14.8cm,clip]{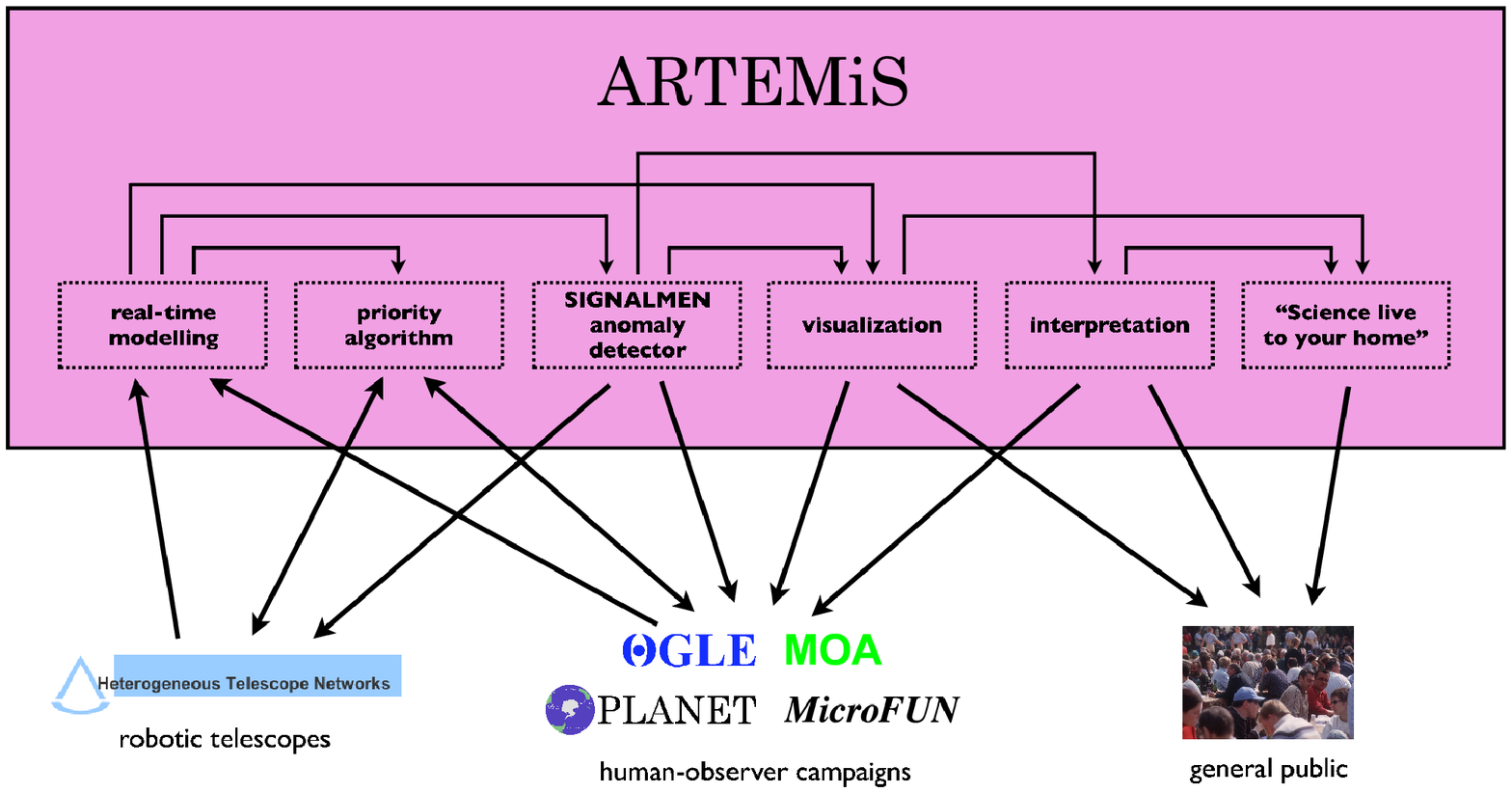}
\end{center}
\caption{ARTEMiS (Automated Robotic Terrestrial Exoplanet Microlensing
Search) and its interactions with the outside world.}
\label{fig:ARTEMISconcept}
\end{figure*}

On 2005 August 10, PLANET/RoboNet, OGLE, and MOA observed a 15~\% deviation
to the light curve of microlensing event OGLE~2005-BLG-390 over about a
day, which was shown to be due to a planet of about 5~Earth masses,
orbiting a star with $0.2~M_\odot$ at 3~\mbox{AU} with a period of 10~years, where all these values are uncertain to a factor of two
(Beaulieu et al. 2006).
An Earth-mass planet in the same spot would still have caused a
signal amplitude of 3\,\% and duration of $\sim\,12$~h. If one
assumes photometric uncertainties of $\sim\,1$\,\%, the discovery
of such a planet would only have been possible if the standard
follow-up sampling of 2~h had been replaced by high-cadence
(10--15~min) anomaly monitoring triggered upon the first suspicion of a deviation.
Real-time photometry and a prompt response from the 
telescopes allow the {\sc SIGNALMEN} anomaly detector to identify
ongoing anomalies by successively requesting further observations 
until an anomaly can be confirmed or rejected with the 
required significance.
By triggering on residuals whose absolute value is among the 
largest 5\,\% of all data for the respective site (such trigger points being marked
by arrows in Fig.~1), and eliminating the effect
of outliers by means of robust-fitting techniques, {\sc SIGNALMEN} 
carefully addresses the fact that reported photometric error bars
frequently do not properly represent the true uncertainties and in
general do not follow a Gaussian distribution.

The giant source star ($R_\star \sim 9.6~R_\odot$) that was observed
in OGLE~2005-BLG-390 yielded a larger probability to detect a 
planetary signal and increased its duration, but reduced its amplitude
as compared to a main-sequence star. However, the {\sc SIGNALMEN}
anomaly detector would also allow to reveal an Earth-mass planet from
a 5\,\% deviation
if the observed source is a main-sequence star ($R_\star \sim 1.2~R_\odot$), provided that exposure times are chosen long enough for
achieving a photometric accuracy of 1--2\,\%. 
While the microlensing searches face a reliable chance of first detecting
an extra-solar planet of Earth mass, the detection of 
planets with masses as small as 0.1~$M_\oplus$ is challenging both by
means of the short signal duration and the tiny probability for 
signals of appropriate amplitudes to occur, but possible in principle
(Dominik et al.~2007).

\section{The ARTEMiS concept (2008+)}

The detection of a significant number of terrestrial extra-solar planets
in reasonable time requires a well-coordinated effort involving a network
of either 2m or 4 $\times$ 1m telescopes at its sites and further smaller telescopes, in order to ensure that sufficient data can be acquired on an ongoing planetary anomaly, which could require that observations are being scheduled at a suitable site within 10~min of an alert by the SIGNALMEN anomaly detector. Comprised of the Liverpool Telescope (LT), the Faulkes Telescope North (FTN), and the Faulkes Telescope South (FTS), RoboNet-1.0 constituted the prototype of such a network, with the
two Faulkes telescopes now having been acquired by Las Cumbres Observatory, and further such instruments can be expected to be deployed over the next three years.

Acting as an expert system, ARTEMiS will determine the optimal target to be observed by any site at any given time by means of real-time modelling
of real-time data released by the microlensing observing campaigns and
subsequent assessment by a priority algorithm and the SIGNALMEN anomaly
detector (see Fig.~2). With a flexible strategy catering for parameters that allow observing sites to define their preferences, tailored target recommendations matching specific strategic goals are provided.

For scheduling its observations, RoboNet-1.0 already used the novel software architecture developed by the \mbox{eSTAR} (e-Science Telescopes for Astronomical Research) project (Steele et al.~2002), which
builds a virtual meta-network between existing proprietary robotic-telescope networks providing a uniform interface built upon a multi-agent contract model (Allan, Naylor \& Saunders 2006). 
The embedding of ARTEMiS into eSTAR will open a direct way of communication with robotic telescopes in the HTN (Heterogeneous Telescope Networks) consortium. Moreover, using the standard adopted by the IVOA (International Virtual Observatory Alliance) for representing, transmitting, publishing, and archiving the discovery of a transient celestial event
(White et al.~2006), 
different levels of {\sc SIGNALMEN} alerts on potential or actual anomalies, defining the switch-over from follow-up to anomaly monitoring mode, will be distributed as Virtual Observatory Events (VOEvents).

Nevertheless, ARTEMiS also accounts for the fact that many of the current
microlensing campaigns still rely on human observers. For these,
additional means of communication will be established, such as
interactive webpages and SIGNALMEN alerts circulated as e-mail
or SMS. In particular,
ARTEMiS will offer up-to-the minute visualization of incoming data 
and model light curves, using the system that was previously operated
for PLANET, and provide current information about ongoing anomalies.
The real-time visualization and interpretation not only allows to
link up with professional and amateur astronomers around the world, but even offers an opportunity
to communicate forefront research in progress to the general public 
as ``Science live to your home''.

\end{document}